# Collective atom phase-controlled noise-free photon echoes using double rephasing and optical locking

Byoung S. Ham
Center for Photon Information Processing, School of Electrical Engineering, Inha University
253 Yonghyun-dong, Nam-gu, Incheon 402-751, S. Korea
bham@inha.ac.kr

**Abstract:** Using collective atom phase control a population inversion-free photon echo scheme has been studied for quantum memory applications. For the inversion-free photon echoes a double rephasing method is combined with optical locking, where the optical locking coherently controls each atom phase excited by weak data pulses. As a result, intensity flattened, normally ordered, storage time-extended, and noise-free photon echoes are obtained.

Utilizing optical rephasing in an inhomogeneously broadened optical medium, photon echoes have been studied for multimode and ultrawide photon storage capability in a collective atom (or ion) ensemble [1]. In the rephasing process, however, population inversion is inevitable, where photon echoes can be amplified [2]. This gain process of photon echoes is crucial for all-optical feedback systems such as associative optical memories, where more than 100% echo efficiency is required. However, such an inversion scheme must be avoided for quantum memories, even though echo gain has been minimally observed due to the echo reabsorption process in the propagation direction. In general photon echo efficiency is as low as 1% in most solid media [3].

For quantum memory applications, photon echoes have been modified to solve the population inversion problem and to increase the extremely low retrieval efficiency [4-10]. In these methods, however, trade-off between finesse and absorption [4-7] or usage of paired electrodes [8-10] may deteriorate quantum fidelity or limit full bandwidth usage. Moreover, the observed photon storage time is much shorter than a millisecond, where at least one second is required for long-distance quantum communications utilizing quantum repeaters [11]. Thus, any method to simultaneously solve the population inversion problem and the short photon storage time is urgently needed. Here a simple method to satisfy the required conditions of inversion-free and extended photon storage time is presented for quantum memory applications.

The photon echo storage time is limited by the optical phase decay time of the medium, which is much shorter than a millisecond in most rare-earth doped solids [3]. Recently photon storage time extension has been experimentally demonstrated using optical locking [12], where optical locking relates with coherence conversion between optical and spin states via coherent population transfer. To overcome extremely low photon echo efficiency, a backward propagation scheme using optical locking has been applied for fifteen fold-enhanced echo efficiency [12]. A gradient echo scheme also greatly benefits to enhanced echo efficiency [8-10].

Recently a double rephasing technique has been proposed to solve the rephasing pulse-caused population inversion [13]. In ref. 13, however, the double rephasing scheme accompanies conventional photon echoes open to stimulated gain or spontaneous emission process. Subsequently, a French group presented a silenced echo scheme using the double rephasing technique, where the first echo generation is severely reduced [14]. An Australian group applied the double rephasing scheme to gradient echoes based on paired electrodes [15], where quantum noise is silenced. These methods are still limited, however, by short storage time or coherence loss.

In Fig. 1, the optical locking scheme with a B1 and B2 control pulse pair is presented to show collective atom phase control. The optical locking was originally proposed for resonant Raman echoes to extend photon storage up to spin population decay time [16]. For Fig. 1, a three-level lambda-type optical system is concerned, whose energy levels are composed of |1>, |2>, and |3>. The data (D) and rephasing (R) pulses are resonant to the transition of |1>−|3>, while the optical locking pulses, B1 and B2 are resonant to the transition of |2>−|3>, where |3> is the excited state. The function of the optical locking pulse B1 is to halt the rephasing process initiated by R by transferring the excited atoms into the robust spin state |2>, where the spin decay time is much longer than the optical counterpart [17]. Returning the transferred atoms to the original state |3> by B2 with atom phase compensation, the halted rephasing process resumes and generates a conventional photon echo signal under the population inversion as shown in Fig. 1(a) (see the dotted red circle).

Figure 1(b) shows coherence evolution of the inhomogeneously broadened atom ensemble, where 161 divisions are made for time-dependent density matrix calculations in the 800 kHz Gaussian distributed atom ensemble (see the red curve in Fig. 1(c) for Re$\rho_{13}$; FWHM: 340 kHz): Figure 1(a) is the sum of all spectral components. In Figure 1(c), one detuned atom group is selected from Fig. 1(b) to address the rephasing process via optical locking resulting in photon storage-time extension. Although this extension is limited by the spin dephasing process in the experiment [12], all decay rates are set at zero for visualization purposes. Each coherent population transfer between states |3> and |2> accompanies a $\pi/2$ phase shift [17]. Thus, the optical locking condition becomes a $2n\pi$ for the phase shift, where n is an integer [17]. Figures 1(c) and 1(d) represent atom phase control by adjusting the B2 pulse area, where the $\pi-7\pi$ (or $\pi-3\pi$) B1−B2 pulse sequence results in no change to the coherence rephased by R1. As shown in Fig. 1(d) the use of an identical pulse set of B1−B2 results in a $\pi$ phase shift prohibiting photon echo generation due to the absorptive characteristics (see the dotted arrow): $\exp(i\delta t) \rightarrow \exp(-i\delta t)\exp(i\pi) = -\exp(-i\delta t)$. Here, the storage time extension determined by B2 delay can be up to in the order of 100 milliseconds if an external magnetic field gradient is applied [18,19].

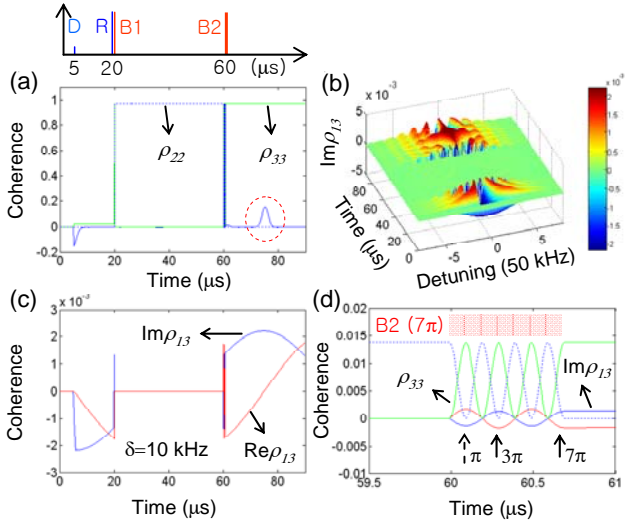

Figure 1. Schematics of atom phase control in photon echoes. (a) Optically locked photon echo. Inset: pulse sequence; D (data): $0.1\pi$ in pulse area: R (rephasing), B1, and B2: $\pi$, $\pi$, and $3\pi$ in pulse area. (b) Coherence versus time and detuning. (c) $\delta$–detuned atom phase evolution. (d) Extended figure of (c). FWHM of the Gaussian distributed atoms is 340 kHz. All decay rates are set at zero. D pulse duration is 1 µs, while the pulse duration of R and B1 is 0.1 µs. For (d) and (d) $\delta$=10 kHz.

In Fig. 2, we discuss how to manipulate the atom phase for inversion-free photon echoes. Figure 2 shows numerical calculations for the inversion-free photon echoes. The insets on top of Figs. 2(a) and 2(b) represent the corresponding optical pulse timing. This pulse sequence is achieved by adding the second rephasing pulse R2 to that of Fig. 1 with pulse area adjustment of B2 (see the dotted arrow in Fig. 1(d)). Figure 2(a) shows echoes with (E1) and without (E2) population inversion, where the red line denotes excited state population $\rho_{33}$ (see Fig. 2(b)). Unlike the photon echo in Fig. 1(a), E1 in Fig. 2(a) is absorptive due to the use of $\pi$ B2 pulse as discussed in Fig. 1(d), resulting in no echo generation. Echo absorption is also prohibited due to the population inversion. Thus, echo E1 does not alter the system coherence representing a silenced echo. Figure 2(b) shows atom population change, where E2 is under no population inversion by R2. Thus, the echo E2 must be a complete retrieval of the data D without any spontaneous or stimulated emission noise. Here, the E2 evolution direction progresses forward as D-excited coherence due to the double rephasing by R1 and R2 (discussed in Fig. 3).

Figures 2(c) and 2(d) represent all atoms' coherence evolutions for absorption (Im$\rho_{13}$) and dispersion (Re$\rho_{13}$), respectively. As shown, the atom phase is reversed by R1 at t=20 µs and reversed again by R2 at t=90 µs.

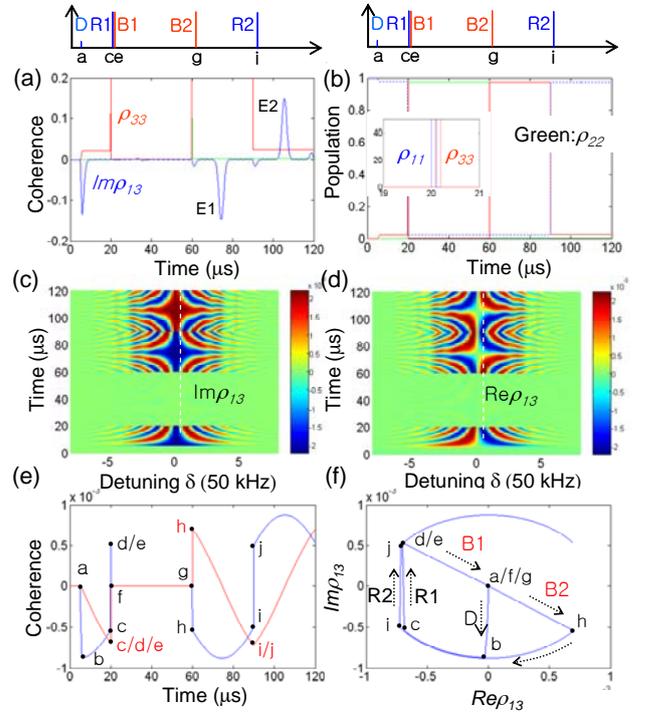

Figure 2. (a) Numerical calculations for inversion-free echo generation using double rephasing and optical locking. Each pulse area of R1, R2, B1, and B2 is $\pi$. The pulse area of D is $0.1\pi$. The pulse duration of D is 1 µs. Pulse duration of R1, R2, B1 and B2 is 100 ns, respectively. (b) Population versus time. (c) and (d) Coherence as a function of time and detuning of (a). (e) Coherence for a detuned atom by $\delta$=10 kHz. Each letter indicates pulse timing of pulses in the inset of Fig. 1. (f) Bloch vector notation for a detuned atom of (e). FWHM of the Gaussian distributed atoms is 340 kHz. All decay rates are set at zero.

Figures 2(e) and 2(f) show details of coherence evolution for a detuned atom group ($\delta$=10 kHz) picked up in Figs. 2(c) and 2(d) (see the dashed lines). By data pulse D (a-b) optical coherence is excited due to absorption, and then freely evolves (b-c) at the speed determined by the detuning δ: exp(iδt). The first rephasing pulse R1 (c-d) swaps populations between states |1> and |3>, and gives a π phase shift to each Imρ$_{13}$ and Reρ$_{13}$: exp(iδt) → exp(−iδt). As a result, the evolution direction is reversed, representing time reversal process. This time reversal is the key mechanism of photon echo formation. By the optical locking pulse pair B1 (e-f) and B2 (g-h), population in the excited state |3> experiences a round-trip travel to state |2> via double population transfer from |3> to |2> and back to |3>. However, these identical B1-B2 interactions alter the atomic coherence by a π phase shift resulting in absorptive echo E1: exp(−iδt) → exp(−iδt)exp(iπ) = −exp(−iδt). The atom phase must be modified after E1 by corssing over the v-axis (see Fig. 1(f)): −exp(−iδt) → exp(iδt). By the second rephasing pulse R2 (i-j) populations between states |1> and |3> are re-swapped, and another π phase shift is given to the atomic coherence: exp(iδt) → exp(−iδt). This final state indicates emission of photons like a conventional photon echo, but with no population inversion. The coherence evolution direction is, however, forward, resulting in a normally ordered echo sequence (see Fig. 3). Thus, coherent control of the optical locking pulse pair B1 and B2 results in the silenced echo E1 (absorptive under population inversion) and inversion-free echo E2.

Figure 2(f) shows a Bloch vector-like model in a uv plane, where u and v indicate the real and imaginary parts of coherence ρ$_{13}$, respectively. The function of the rephasing pulse (R1 and R2) is to flip over the uv plane along the u-axis. The function of each optical locking pulse B1 or B2 is to give a π/2 phase shift, so that the identical pulse pair of B1 and B2 at each π pulse area results in double flips along both the u- and v-axis as known. Here an additional function of B1 via coherence transfer from state |3> to |2> is to extend storage time dependent upon the spin dephasing time (no spin dephasing is assumed in the calculations).

Figure 3 shows multiple bit storage to show intensity flattened normally ordered photon echoes. The light pulse stream is composed of a, b, 0 and c, where a, b, and c denote "on." As shown in Fig. 3(a), the order of photon echo E1 is reversed, while E2 is in the same order as the data pulses. More importantly, E2 amplitude (or intensity) in the data stream is flat due to exact compensation between γ−dependent coherence decay (see the dotted line fitted for the red echoes) as a function of time and the reversed read-out of E1 by R2:

$$I_{E2} = I_{E1} \exp\left(-\frac{T_{E2}-T_{E1}}{\tau}\right) = I_0 \exp\left(-\frac{T_{E2}-T_D}{\tau}-\alpha\right), \quad (1)$$
$$I_{E1} = I_0 \exp\left(-\frac{T_{E1}-T_D}{\tau}-\alpha\right),$$

where $I_0$ is the input data intensity, and $\alpha$ is a decay factor given by the spin dephasing. Here ($T_{E2}-T_D$) is the same for all data pulses (a, b, and c), because echo E2 has the same order as the data. In a slow decaying medium like a rare earth $Pr^{3+}$ doped $Y_2SiO_5$, ($T_{E2}-T_D$) can be made smaller compared with τ, where τ is the optical homogeneous decay time. In this case, the optical locking time T ($T=T_{B2}-T_{B1}$) becomes a dominant factor to determine the storage time if $\alpha$ is a small number. Figure 3(b) represents the Bloch vector model for symmetrically δ−detuned atoms, where echo E2 is formed when two coherence evolutions meet together (see the arrow mark in the upper center): $T_{E2}=2T_{R2}-(T_{R1}+T_D+T)$. Here, the functions of R1 and R2, and B1 and B2 are depicted by the dotted arrows as explained in Fig. 2(d).

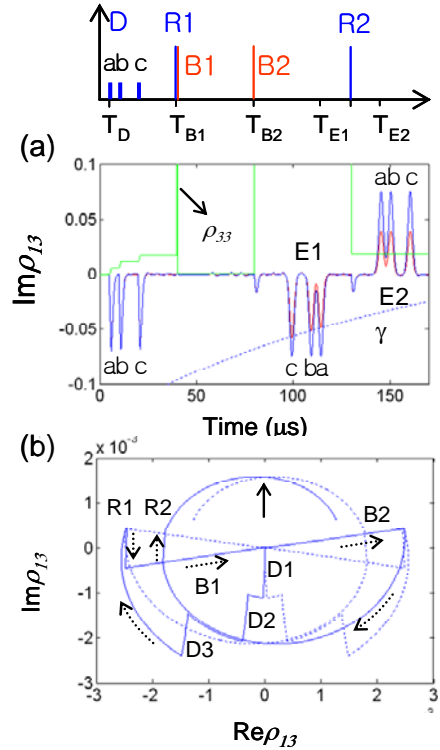

Figure 3. (a) Numerical calculations for multiple inversion-free echoes. Dotted line: γ−dependent best-fit curve for the red echoes. For blue, $\Gamma_{31}=\Gamma_{32}=0$ kHz; $\gamma_{31}=\gamma_{32}=0$ kHz. For red, $\Gamma_{31}=\Gamma_{32}=0$ kHz; $\gamma_{31}=\gamma_{32}=2$ kHz. (b) Bloch vector model of the red in (a) for δ=10 kHz (solid) and δ=−10 kHz (dotted). All other parameters are the same as in Fig. 2. Dotted arrows indicate evolution direction for the solid curve.

In conclusion a population inversion-free photon echo quantum memory scheme was presented by coherent control of the collective atom phase via optical locking. Unlike time reversed conventional two-pulse photon echoes, the present noise-free echo scheme keeps the same order as the data pulses due to double rephasing without population inversion. Based on the optical locking, a backward propagation scheme is an inherent benefit to overcoming the fundamental problem of echo reabsorption in a forward propagation scheme. The flat echo intensity offers a practical benefit to multimode photon storage in the time domain. Finally, photon storage time extension by optical locking via coherent population transfer opens a door to ultralong quantum memories for long distance quantum communications.

This work was supported by the Creative Research Initiative Program (grant no. 2011-0000433) of the Korean Ministry of Education, Science and Technology via the National Research Foundation.

Reference
1. N. A. Kurnit, I. D. Abella, and S. R. Hartmann, "Observation of a photon echo," Phys. Rev. Lett. **13**, 567-570 (1964).
2. C. S. Cornish, W. R. Babbitt, and L. Tsang, "Demonstration of highly efficient photon echoes," Opt. Lett. **25**, 1276-1278 (2000).
3. R. M. Macfarlane and R. M Shelby, "Coherent Transient and Holeburning Spectroscopy of Rare Earth Ions in Solids," in Spectroscopy of Solids Containing Rare Earth Ions, Kaplyanskii, A. & Macfarlene, R. M. eds.(North-Holland, Amsterdam, 1987).
4. H. de Riedmatten, M. Afzelius, M. U. Staudt, C. Simon, and N. Gisin, "Solid-state light-matter interface at the single-photon level," Nature **456**, 773-777 (2008).
5. M. Afzelius, I. Usmani, A. Amari, B. Lauritzen, A. Walther, C. Simon, N. Sangouard, J. Minar, H. de Riedmatten, N. Gisin, and S. Kröll, "Demonstration of atomic frequency comb memory for light with spin-wave storage," Phys. Rev. Lett. **104**, 040503 (2010).
6. I. Usmani, M. Afzelius, H. de Riedmatten, and N. Gisin, "Mapping multiple photonic qubits into and out of one solid-state atmic ensemble," Nature Communi. **1**, 12-19 (2010).
7. C. Clausen, I. Usmani, F. Bussieres, N. Sangouard, M. Afzelius, H. de Riedmatten, and N. Gisin, "Quantum storage of photonic entanglement in a crystal," Nature **469**, 508-511 (2011).
8. G. Hetet, J. J. Longdell, A. L. Alexander, P. K. Lam, and M. J. Sellars, "Electro-Optic quantum memory for light using two-level atoms," Phys. Rev. Lett. **100**, 023601 (2008).
9. B. Hosseini, B. M. Sparkes, G. Sparkes, G.. Hetet, J. J. Longdell, P. K. Lam, and B. C. Buchler, "Coherent optical pulse sequencer for quantum applications," Nature **461**, 241-245 (2009).
10. M. P. Hedges, J. J. Longdell, Y. Li, and M. J. Sellars, "Efficient uantum memory for light," Nature **465**, 1052-1056 (2010).
11. W. Dür, H.-J. Briegel, and P. Zoller, "Quantum repeater," in Lectures on quantum information, D. Bruß, ed. (Wiley-VCH Verlag GmbH & Go., 2007).
12. J. Hahn and B. S. Ham, "Rephasing halted photon echoes using controlled optical deshelving," New J. Phys. **13**, 093011 (2011).
13. B. S. Ham, "Atom phase controlled noise-free photon echoes," arXiv:1101.5480 (2011).
14. V. Damon, M. Bonarota, A. Louchet-Chauvet, T. Chaneliere, and J.-L. Le Gouet, "Revival of silenced echo and quantum memory for light," arXiv:1104.4875v1 (2011).
15. D. L. McAuslan, P. M. Ledingham, W. R. Naylor, S. E. Beavan, M. P. Hedges, M. J. Sellars, and J. J. Longdell, "Photon echo quantum memories in inhomogeneously broadened two-level atoms," arXiv:1104.4134v3 (2011).
16. B. S. Ham, "Ultralong quantum optical storage using reversible inhomogeneous spin ensembles," Nature Photon. **3**, 518-522 (2009).
17. B. S. Ham, "Control of photon storage time using phase locking," Opt. Exp. **18**, 1704-1713 (2010).
18. E. Fraval, M. J. Sellars, and J. J. Longdell, "Method of exteding hyperfine coherence times in Pr3+:Y2SiO5," Phys. Rev. Lett. **92**, 077601 (2004).
19. M. Lovric et al., "Hyperfine characterization and coherence lifetime extension in $Pr:^{3+}:La(WO_4)_3$," arXiv:1107.2274v1 (2011).